\magnification=\magstep1
\baselineskip=20 true pt
\hsize=16 true cm
\vsize=20 true cm
\centerline {\bf Nonequilibrium Phase Transition and 
'Specific-heat' singularity}
\centerline {\bf in the kinetic Ising model: A Monte Carlo Study}
\bigskip
\bigskip
\centerline {Muktish Acharyya$^*$}
\bigskip
\bigskip
\bigskip
\centerline {\it Department of Physics}
\centerline {\it Indian Institute of Science, Bangalore-560012, India}
\centerline {\it and}
\centerline {\it Condensed Matter Theory Unit}
\centerline {\it Jawaharlal Nehru Centre for Advanced Scientific Research}
\centerline {\it Jakkur, Bangalore-560064, India}
\bigskip
\bigskip
\leftline {\bf Abstract:} 
The nonequilibrium phase transition has been studied 
by Monte Carlo simulation in a
ferromagnetically interacting (nearest neighbour)
kinetic Ising model in presence of a sinusoidally oscillating 
magnetic field. 
The ('specific-heat') 
temperature derivative of 
energies (averaged over a full cycle of the oscillating field) 
diverge near the dynamic transition
point.
\bigskip
\bigskip
\bigskip
\bigskip
\bigskip
\leftline {\bf $^*$E-mail:muktish@physics.iisc.ernet.in}
\vfil \eject
\leftline {\bf I. Introduction}

The dynamics of Ising system in presence of an oscillating magnetic field
has been studied extensively [1,3] in the last few years. 
The dynamic hysteretic response [1,3]
and the nonequilibrium dynamic phase transition [2,3] became an interesting 
and important part of
these studies. 
The nonequilibrium dynamic phase transition in 
the kinetic Ising model in presence
of a sinusoidally oscillating magnetic field has been studied first 
by Tome
and Oliviera [2]. They have solved the dynamic mean field equation of motion 
for the average magnetisation of 
the kinetic Ising model in presence of a 
sinusoidally oscillating magnetic field. 
Defining the dynamic order parameter as the time averaged magnetisation
over a full cycle of the oscillating field,
they showed that in the field 
amplitude and temperature plane there exists a phase boundary separating
dynamic ordered and disordered phase. 
Precisely, below the phase boundary the value of the order parameter is
nonzero and it gets zero value above the bounday.
There exists a tricritical point on
the phase boundary line separating the nature (discontinuous/continuous) of
the dynamic transition.

Later, Acharyya and Chakrabarti [3] studied this 
nonequilibrium phase transition
in the kinetic Ising model by extensive Monte Carlo simulations.
They have also studied [3] the
temperature variations of the AC susceptibility components. Their 
important observation was that the imaginary (real) part of AC susceptibility
gives a peak (dip)
at the dynamic transition point, indicating
the thermodynamic nature of this type of
 nonequlibrium dynamic phase transition.

Here, in this paper the time averaged (over a full cycle
of the oscillating field) cooperative energy
and total energy of the system have
 been calculated by Monte Carlo simulation
and studied the temperature
variation of the temperature derivatives ('specific-heat') of those energies.
These 'specific-heat' like quantities are observed to diverge very near the
dynamic transition point (where the dynamic order parameter vanishes).
The paper is organised as
follows: in section II the model and the simulation scheme are
discussed very briefly, the results are reported in section III
and the paper ends with some concluding remarks in section IV.
\bigskip
\bigskip
\leftline {\bf II. Model and Simulation}

A ferromagnetically interacting 
(nearest neighbour) Ising system in presence
of an oscillating magnetic field can be represented by the 
Hamiltonian,
$$H = -\sum_{<ij>} J_{ij} s_i^z s_j^z - h(t) \sum_i s_i^z. $$
Here, $s_i^z (= \pm 1)$ 
represents Ising spin variable and, 
$h(t) = h_0 cos (\omega t)$ is the externally
applied oscillating magnetic field. $h_0$ and $\omega$ are the amplitude
and frequency of the field. For simplicity all $J_{ij} (> 0)$'s are taken
here equal to unity and boundary condition is periodic.

Here, a square lattice of 
linear size $L (= 100)$ has been considered. 
At any finite
temperature $T$ and for a fixed frequency ($\omega$) 
and amplitude ($h_0$) of the
field, the 
dynamics of this system has been studied here by Monte Carlo
simulation using Glauber single spin-flip dynamics. Each lattice site is
updated here sequentially and one such full scan over the lattice is
defined as the time unit here.

The following quantities are calculated, for a fixed values of $T$,
$\omega$ and $h_0$:

(i) The dynamic order parameter,
$Q = (\omega/{2 {\pi}})\oint m(t) dt,$ 
where $m(t)$ is the instanteneous magnetisation.
This is essentially the time ($t$) averaged magnetisation over a full cycle of
the oscillating magnetic field.

(ii) The time averaged (over a full cycle) cooperative energy of the system
$$E_{coop} = -(\omega/{2 {\pi} L^2}) \oint 
\left(\sum_{<ij>} s_i^z s_j^z \right) dt. $$

(iii) The time averaged (over a full cycle) total energy of the system
$$E_{tot} = -(\omega/{2 {\pi} L^2})\oint \left(\sum_{<ij>} s_i^z s_j^z 
+ h(t) \sum_i s_i^z \right)dt. $$

Each data point has been obtained by averaging over 50 different random
Monte Carlo samples.
\vfil \eject
\leftline {\bf III. Results}

The temperature derivatives of $E_{coop}$ and $E_{tot}$ can be called
'specific-heat' for this nonequilibrium problem.
The temperature variations of $Q$ , 
$C_{coop}= {\delta E_{coop}}/ {\delta T}$ and 
$C_{tot}= {\delta E_{tot}}/{\delta T}$ have been studied. 
For a fixed values of
the field amplitude $h_0$, the dynamic order parameter 
$Q$ shows a continuous
(depending upon the value of $h_0$) phase transtion at 
temperature $T_d(h_0)$ [3].
The values of $h_0$ (=0.4 and 0.8)
are chosen here in such a way that $Q$ always undergoes
a continuous transition.
The temperature variations of $Q$, 
$C_{coop}$ and $C_{tot}$ have been shown in Fig. 1.
In this case the frequency ($\omega$) of the field is kept fixed 
($\omega$ = 0.0628). 
From the figure it is clear that the 'specific-heat' s
$C_{coop}$ and $C_{tot}$  diverge near the dynamic phase transition point.
$C_{tot}$ shows more prominent divergence than that showed by $C_{coop}$.
Some of the results have been checked for larger sizes ($L = 200$) and no
significant deviation was observed. The CPU time used here, to generate the
whole set of data in IBM workstation, is approximately 16 hours.

\bigskip
\leftline {\bf IV. Concluding remarks}

In conclusion, the nonequilibrium phase transition in 
the kinetic Ising model has
been studied by Monte Carlo simulation. The 'specific-heat' shows
divergence near the dynamic transition point. 
The divergence of 'specific-heat' near the dynamic transition point is
a distinct signal of phase transition and an indication of the thermodynamic
nature of this class of nonequilibrium phase transition.
The
indication of the
thermodynamic nature of this type of nonequilibrium transition was first 
observed
by Acharyya and Chakrabarti [3] in their study of the temperature
variation of AC
susceptibility. 
This
observation of 'specific-heat' divergence near the transition point
re-confirms,
by another independent way,
the thermodynamic nature of dynamic phase transition. 
A lot of computational effort is required
to establish precisely 
the divergence of 'specific-heat' near the transition point.
The work is in progress in this
front and the details will be reported elsewhere.

It would be quite interesting to know how the
dynamic order parameter goes to zero and specific heat diverges 
as one approaches the transition point.
The universality class of this type of
nonequilibrium transition, is not yet known.

\bigskip
\leftline {\bf Acknowledgements:} Author is grateful to JNCASR for financial
help and SERC, Bangalore for computational facilities. 
Author is thankful to
R Lahiri for a careful reading of the manuscript.
\bigskip
\centerline {\bf References:}

\item{[1]} M. Rao, H. R. Krishnamurthy and R. Pandit, Phys. Rev. B
{\bf 42} (1990) 856; P. B. Thomas and D. Dhar, J. Phys. A: Math Gen
{\bf 26} (1993) 3973; S. Sengupta, Y. Marathe and S. Puri, Phys. Rev.
B {\bf 45} (1990) 4251; M. Acharyya and B. K. Chakrabarti, in {\it
Annual Reviews in Computational Physics}, Ed. D. Stauffer, 
World-Scientific (Singapore), Vol 1
(1994) pp 107
\item{[2]} T. Tome and M. J. de Oliviera, Phys. Rev. A {\bf 41} (1990) 4251
\item{[3]} M. Acharyya and B. K. Chakrabarti, Phys. Rev. B {\bf 52} (1995)
6550; see also M. Acharyya and B. K. Chakrabarti, J. Mag Mag Mat {\bf 136}
(1994) L29
\vfil \eject
\centerline {\bf Figure Caption}

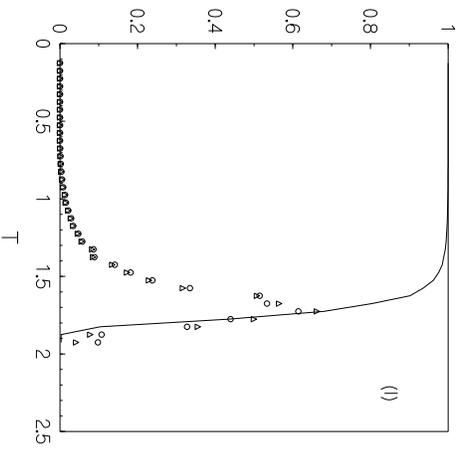
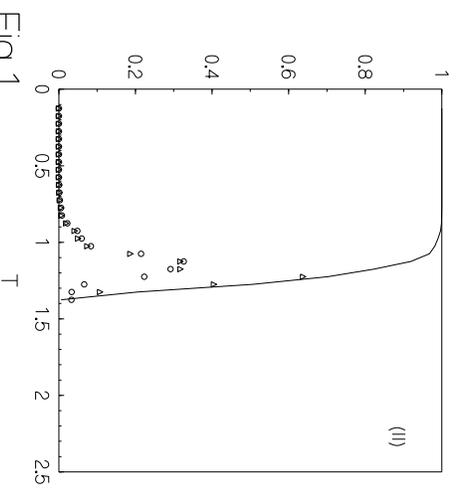
Fig.1. Temperature variation of $Q$ (solid line), $C_{coop}$ (circle)
and $C_{tot}$ (triangle) for fixed $\omega$ (= 0.0628) and two different
values of the field amplitude ($h_0$): (I) $h_0$ = 0.4
and (II) $h_0$ = 0.8. $L$ = 100 here.
\end